\documentclass[preprint,preprintnumbers,amsmath,amssymb,nofootinbib]{revtex4}

\usepackage{graphicx}
\usepackage{dcolumn}
\usepackage{bm}

\begin{document}


\title{Inhomogeneous magnetic field in AdS/CFT superconductor}

\author{Wen-Yu Wen}
\email{steve.wen@gmail.com}
\affiliation{Department of Physics and Center for Theoretical Sciences\\ 
National Taiwan University, Taipei 106, Taiwan}


\begin{abstract}
We study the holographically dual description of superconductor in $(2+1)$-dimensions in the presence of inhomogeneous magnetic field and observe that there exists type I and type II superconductor.  A new feature of type changing is observed for type I superconductor near critical temperature.
\end{abstract}


\maketitle

\section{Introduction}
The holographic correspondence between a gravitational theory and a quantum field theory, first emerged under the AdS/CFT correspondence\cite{Maldacena:1997re}, has been proved useful to study various aspects of nuclear physics such as RHIC and condensed matter phenomena, particularly in those recent studies \cite{Herzog:2007ij,Hartnoll:2007ih,Hartnoll:2007ip,Hartnoll:2008hs,Minic:2008an}.

In the papers \cite{Gubser:2005ih,Gubser:2008px}, the author proposed a gravity model in which abelien symmetry of Higgs is spontaneously broken by the existence of black hole.  This mechanism was recently used to study non-abelian gauge condensate\cite{Gubser:2008zu} and to our interests, the model of superconductivity where critical temperature was observed\cite{Hartnoll:2008vx}.  This model was later studied in the presence of magnetic field and critical magnetic field was also observed\cite{Nakano:2008xc}.  With electromagnetic field fully back-reacted, the condensate was observed to be localized in one dimension implying the Meissner effect\cite{Albash:2008eh}.

In this note, we would like to extend our previous work\cite{Nakano:2008xc} to the construction of gravitational dual model of superconductor in the presence of inhomogeneous magnetic field.

To implement an inhomogeneous magnetic field at finite temperature, we prefer to switch on the electromagnetic field in the GL matter sector without back reaction to the gravity sector.  This is due to the construction of non-planar RN black hole is not available so far.  We will show that with this simplified picture, it is enough to obtain the expected type I and II superconductor as in conventional GL model but with a novel feature of type changing.  At the end, we will improve our model to compromise with the desired inhomogeneous magnetic field by including partial back reaction.

\section{The model with inhomogeneous magnetic field}
Several important unconventional superconductors, such as the cuprates and organics, are layered in structure and interesting physics can be captured by studying a $(2+1)$ dimensional system.  We have been successful in building up a gravity model (in coupled with other matter fields) in $(3+1)$ dimensions which is holographically dual to the desired planar system which develops superconductivity below critical temperature and critical magnetic field\cite{Nakano:2008xc}.  We now start with a model composed of the gravity sector and the matter sector.  The gravity sector is given by the AdS Schwarz-Schild black hole in $AdS_4$\cite{Romans:1991nq},
\begin{eqnarray}
&&ds^2=-f(r)dt^2+\frac{dr^2}{f(r)}+r^2(dx^2+dy^2),\\\nonumber\\
&&f(r)=\frac{r^2}{L^2}-\frac{M}{r}.
\end{eqnarray}
Then the temperature is determined via the relation 
\begin{equation}
T=\frac{3M^{1/3}}{4\pi L^{4/3}}.
\end{equation}

For the matter sector, we will use the Ginzburg-Landau (GL) action for a Maxwell field and a charged complex scalar with $m^2L^2=-2$, which does not back react on the metric \cite{Gubser:2008px,Hartnoll:2008vx}, 
\begin{equation}
e^{-1}{\cal L}_m=-\frac{1}{4}F^{ab}F_{ab}+\frac{2}{L^2}|\Psi|^2-\frac{g}{2L^4}|\Psi|^4-|\partial \Psi-iA\Psi|^2.
\end{equation}

We make the following assumption for static solution: $A_t=\Phi(r), A_y=a(x), A_r=A_x=0$ as well as $\Psi(r,x)=\psi(r)s(x)$.  Then we need to solve four coupled second order differential equations
\begin{eqnarray}
&&\label{eq:s}\ddot{s}-a^2 s-\frac{g}{L^4}r^2\psi^2 s^3=-k^2 s,\\
&&\label{eq:psi}\psi''+(\frac{2}{r}+\frac{f'}{f})\psi'+\frac{1}{f}(\frac{2}{L^2}+\frac{\Phi^2}{f}-\frac{k^2}{r^2})\psi=0,\\
&&\label{eq:phi}\frac{1}{r^2} (r^2\Phi')'-\frac{2}{f} (\psi s)^2\Phi=0,\\
&&\label{eq:a}\kappa^2\ddot{a}-2r^2(\psi s)^2a=0,
\end{eqnarray}
Here we have used the prime to denote derivative with respect to $r$ and the dot to $x$.  The first two equations, derived from variation with respect to $\Psi$, have been obtained and solved in \cite{Albash:2008eh} for $g=0$; however the last two, derived from variation with respect to $A_{\mu}$, cannot not be consistently solved due to our ignorance of new degrees of freedom introduced via spatial dependence.  We will come back to this issue in the next section.  Nevertheless, equation (\ref{eq:s}) and (\ref{eq:a}) are just enough for us to study the spatial variation of order parameter $s(x)$ in the presence of inhomogeneous magnetic field $H=\kappa \dot{a}(x)$.  We first recall the asymptotic behaviour of $\psi(r)$ at the boundary, up to the third order: 
\begin{equation}
\psi=\frac{\psi_1}{r}+\frac{\psi_2}{r^2}+\frac{\psi_3}{r^3}+{\cal O}(r^{-4}).
\end{equation}
In this note, we will only focus on the case where $\psi_1\neq 0$ and $g\neq 0$.  For convenience, we may define new variables: 
\begin{equation}
\tilde{g}=\frac{g\psi_1^2}{L^4},\qquad \tilde{\kappa}=\frac{\kappa}{\sqrt{2}\psi_1}.  
\end{equation}
For $k^2=1$, we obtain the desired equations at $r\to \infty$:
\begin{eqnarray}
&&\label{eq:a1}\tilde{\kappa}^2 \ddot{a}=s^2 a,\\
&&\label{eq:s1}-\ddot{s}+a^2 s = s - \tilde{g}s^3.
\end{eqnarray}
There are two obvious solutions for constant $\tilde{H}$: $s=0$ for $\tilde{H}\neq 0$ and $s=1/\sqrt{\tilde{g}}$ for  $\tilde{H}=0$.  We expect some nontrivial solution interpolating between these two constant values.  Indeed, it is well known that the above equations describe a type I superconductor for $\tilde{\kappa}\ll 1$ and type II for $\tilde{\kappa}\gg 1$.  We plot the single vortex solution in type II superconductor in the Figure 1.  One can also expect that vortex lattice may form in the type II case with periodic boundary condition on $s(x)$\cite{Abrikosov:1956sx}. A new feature in this model is that $\tilde{\kappa}$ depends on temperature implicitly in the following way:  recall that near critical temperature\cite{Hartnoll:2008vx},  
\begin{equation}
\psi_1 = \frac{<{\cal O}_1>}{\sqrt{2}} \propto (1-\frac{T}{T_c})^{1/2}.
\end{equation}
Therefore even for a type I superconductor with small $\tilde{\kappa}$ at $T\ll T_c$, as $T_c$ is approached it will eventually change to type II due to arbitrary small $\psi_1$.

\begin{figure}\label{fig_1}
\includegraphics[width=0.45\textwidth]{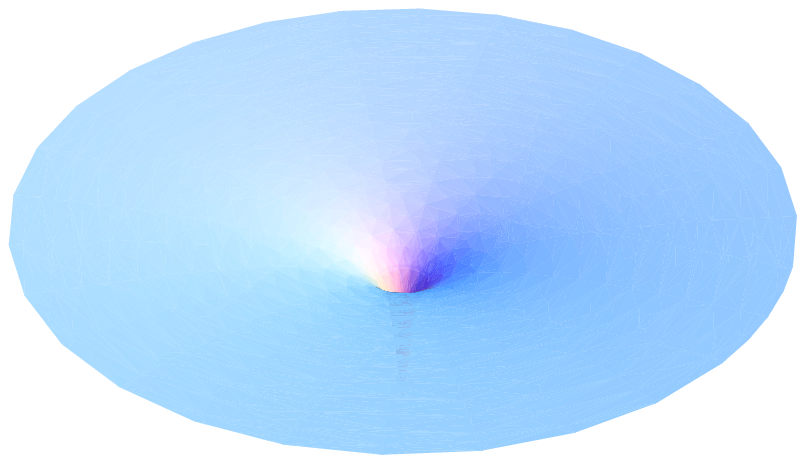}
\includegraphics[width=0.45\textwidth]{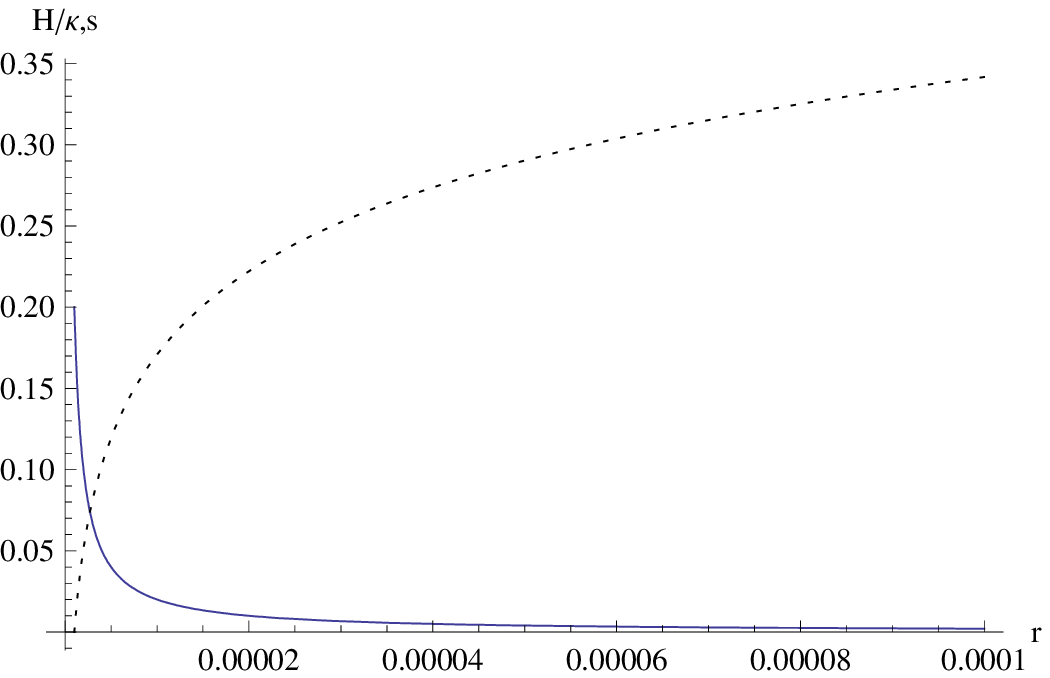}
\caption{To the left, a vortex forms in a type II superconductor where order parameter drops to zero at the center.  To the right is the plot of magnetic field $H/\kappa$ (solid curve) and order parameter $s$ (dotted curve) near the vortex center.  It shows that magnetic flux penetrates through the center of vortex.}
\end{figure}

\section{Back reaction}
We have learnt that both $\Phi(r)$ and $a(x)$ take specific forms in order to be a solution to the Reissner-Nordstr\"{o}m black hole\cite{Romans:1991nq}.  However, we also learn from previous discussion that we need more general ansatz and compatible back-reacted metric in order to study its spatial variation in presence of inhomogeneous magnetic field.  One example is given by solving equation (\ref{eq:a}) providing there is partial back reaction on the metric due to $a(x)$ and $s(x)$ given by equations (\ref{eq:a1}) and (\ref{eq:s1}).  This gives rise to correction to the metric\footnote{We intend not to back react to the equation (\ref{eq:phi}) since it will incur $x$-dependence on function $f$ and break planar symmetry of black hole solution.  Moreover this may spoil the desired solution to the equation (\ref{eq:psi}).}:
\begin{equation}
g_{xx}=g_{yy}=r^2(1-\frac{2\psi_2}{\psi_1r}+\frac{3\psi_2/\psi_1-2\psi_3}{\psi_1r^2}+{\cal{O}}(r^{-3})).
\end{equation}
We remark that with this correction, it looks better for the equation (\ref{eq:s}) because its $r$-dependence drops off thanks to $r^2\psi(r)^2\to g_{xx}\psi(r)^2 = 1$.  However, one has to be warned that this correction might be too naive since more correction terms from several iterations have to be worked out in order to check the convergence of correction and to be consistent with the fully back-reacted Einstein equation if that is desired.  Nevertheless, this correction might be still suitable for large $r$ and slow-varying $s(x)$.

 
\section{Discussion}
In this note, we have considered a hybrid model for AdS/CFT superconductors in the presence of inhomogeneous magnetic field.  Several comments are in order: At first, a magnetic field is provided in the matter sector as a probe, independent of the gravity sector.  The excuse is our lacking knowledge of non-planar black hole solution.  Nevertheless this is sufficient to observe the expected different types of superconductors.  Secondly, the matter sector has no back reaction to the gravity sector, therefore the equation of motion for total Lagrangian is not satisfied.  Although at the end we have included some naive correction mainly contributed from spatial dependence, it might be still interesting to investigate a fully back-reacted action which could be derived from some higher-dimensional theory such as String theory or M-theory.
Thirdly, we have relaxed the ansatz of planar symmetry in order to introduce an inhomogeneous magnetic field.  We were able to construct the solutions for both type I and II superconductors for the choice of $\psi_1\neq 0$ and $k=1$.  A novel feature for type I superconductor is the type changing near $T_c$.  It is also interesting to study the case with vanishing $\psi_1$ and other $k$'s, but we will defer it to the future.  At last, this construction is a tractable phenomenological model of strongly coupled system which may capture some physics of unconventional superconductors.  Though we do not see fermionic degree of freedom from this macroscopic construction, the complex scalar, serving as ordering parameter, seems sufficient to explain such a critical phenomenon as good as the original GL theory.  In order to pursue a microscopic model along this line of reasoning, one may still need to understand better how to realize underlying fermionic degree of freedom in the context of AdS/CFT correspondence, we expect that recently developed correspondence between gravity and non-relativistic conformal theory\cite{Son:2008ye,Balasubramanian:2008dm} may shed some light in this direction.

\begin{acknowledgments}
The author would like to thank invitation of NTNU to present this work in its final stage.  I am grateful to useful discussion with Kazuyuki Furuuchi, Takayuki Hirayama, Pei-Ming Ho, Hsien-Chung Kao, Feng-Li Lin, Eiji Nakano and Dan Tomino.  The author is partially supported by the Taiwan's National Science Council and National Center for Theoretical Sciences under Grant No. NSC96-2811-M-002-018 and NSC97-2119-M-002-001.
\end{acknowledgments}


\bibliography{apssamp}

\begin{thebibliography}{99}

\bibitem{Maldacena:1997re}
  J.~M.~Maldacena,
  ``The large N limit of superconformal field theories and supergravity,''
  Adv.\ Theor.\ Math.\ Phys.\  {\bf 2}, 231 (1998)
  [Int.\ J.\ Theor.\ Phys.\  {\bf 38}, 1113 (1999)]
  [arXiv:hep-th/9711200].

\bibitem{Herzog:2007ij}
  C.~P.~Herzog, P.~Kovtun, S.~Sachdev and D.~T.~Son,
  ``Quantum critical transport, duality, and M-theory,''
  Phys.\ Rev.\  D {\bf 75}, 085020 (2007)
  [arXiv:hep-th/0701036].

\bibitem{Hartnoll:2007ih}
  S.~A.~Hartnoll, P.~K.~Kovtun, M.~Muller and S.~Sachdev,
  ``Theory of the Nernst effect near quantum phase transitions in condensed
  Phys.\ Rev.\  B {\bf 76}, 144502 (2007)
  [arXiv:0706.3215 [cond-mat.str-el]].

\bibitem{Hartnoll:2007ip}
  S.~A.~Hartnoll and C.~P.~Herzog,
  ``Ohm's Law at strong coupling: S duality and the cyclotron resonance,''
  Phys.\ Rev.\  D {\bf 76}, 106012 (2007)
  [arXiv:0706.3228 [hep-th]].

\bibitem{Hartnoll:2008hs}
  S.~A.~Hartnoll and C.~P.~Herzog,
  ``Impure AdS/CFT,''
  arXiv:0801.1693 [hep-th].

\bibitem{Minic:2008an}
  D.~Minic and J.~J.~Heremans,
  ``High Temperature Superconductivity and Effective Gravity,''
  arXiv:0804.2880 [hep-th].

\bibitem{Gubser:2005ih}
  S.~S.~Gubser,
  ``Phase transitions near black hole horizons,''
  Class.\ Quant.\ Grav.\  {\bf 22}, 5121 (2005)
  [arXiv:hep-th/0505189].

\bibitem{Gubser:2008px}
  S.~S.~Gubser,
  ``Breaking an Abelian gauge symmetry near a black hole horizon,''
  arXiv:0801.2977 [hep-th].

\bibitem{Gubser:2008zu}
  S.~S.~Gubser,
  ``Colorful horizons with charge in anti-de Sitter space,''
  arXiv:0803.3483 [hep-th].
   
\bibitem{Hartnoll:2008vx}
  S.~A.~Hartnoll, C.~P.~Herzog and G.~T.~Horowitz,
  ``Building an AdS/CFT superconductor,''
  arXiv:0803.3295 [hep-th].
  

\bibitem{Nakano:2008xc}
  E.~Nakano and W.~Y.~Wen,
  ``Critical magnetic field in AdS/CFT superconductor,''
  arXiv:0804.3180 [hep-th].

\bibitem{Albash:2008eh}
  T.~Albash and C.~V.~Johnson,
  ``A Holographic Superconductor in an External Magnetic Field,''
  arXiv:0804.3466 [hep-th].



\bibitem{Romans:1991nq}
  L.~J.~Romans,
  ``Supersymmetric, cold and lukewarm black holes in cosmological
  Einstein-Maxwell theory,''
  Nucl.\ Phys.\  B {\bf 383}, 395 (1992)
  [arXiv:hep-th/9203018].


\bibitem{Abrikosov:1956sx}
  A.~A.~Abrikosov,
  ``On the Magnetic properties of superconductors of the second group,''
  Sov.\ Phys.\ JETP {\bf 5} (1957) 1174
  [Zh.\ Eksp.\ Teor.\ Fiz.\  {\bf 32} (1957) 1442].




\bibitem{Son:2008ye}
  D.~T.~Son,
  ``Toward an AdS/cold atoms correspondence: a geometric realization of the
  Schroedinger symmetry,''
  arXiv:0804.3972 [hep-th].

\bibitem{Balasubramanian:2008dm}
  K.~Balasubramanian and J.~McGreevy,
  ``Gravity duals for non-relativistic CFTs,''
  arXiv:0804.4053 [hep-th].



\end{thebibliography}

\end{document}